\documentclass[pra,twocolumn]{revtex4}
\usepackage{mathtools}
\usepackage{amsmath, amsthm, amssymb}

\newcommand{\sfrac}[2]{\ensuremath{\mathchoice{\tfrac{#1}{#2}}{\tfrac{#1}{#2}}{\frac{#1}{#2}}{\frac{#1}{#2}}}}
\newcommand{\half}{\sfrac{1}{2}}
\newcommand{\Left}{\mathopen{}\mathclose\bgroup\left}
\newcommand{\Right}{\aftergroup\egroup\right}

\usepackage[braket, qm]{qcircuit}
\usepackage{amsmath}

\usepackage{pgfplots}

\protected\def\[#1\]{\begin{align}#1\end{align}}

\renewcommand{\Im}{\operatorname{Im}}

\begin{document}

\title{Gradients of parameterized quantum gates using the parameter-shift rule and gate decomposition}
\author{Gavin E.\ Crooks}
\email{gec@threeplusone.com}
\affiliation{ 
California Institute of Technology,
Pasadena, CA 91125, USA}
\affiliation{Berkeley Institute for Theoretical Sciences, Berkeley, CA 94706, USA}
\begin{abstract}
The parameter-shift rule is an approach to measuring gradients of quantum circuits with respect to their parameters, which does not require ancilla qubits or controlled operations.  Here, I discuss applying this approach to a wider range of parameterize quantum gates by decomposing gates into a product of standard gates, each of which is parameter-shift rule differentiable.
\end{abstract}
\maketitle

\paragraph*{Introduction -}

The parameter-shift rule is a promising approach to evaluating gradients of parameterized quantum circuits on quantum hardware~\cite{Mitarai2018a,Li2017a,Schuld2019a,Bergholm2018a,Vidal2018a}. 
Suppose we have some objective function $f(\theta)$ of a quantum circuit,
\[
\label{function}
f(\theta) &= \langle \psi| \  U_G^\dagger(\theta)\ A\  U_G(\theta)\ |\psi\rangle 
\]
where the parameterized gate is
\[
\label{gategen}
U_G(\theta) = e^{-i a \theta G} \ .
\]
Here, $G$ is the Hermitian generator of the gate and $a$ is a real constant.
The circuit can feature arbitrary unitaries before and after the gate of interest, but for now we will absorb those dynamics into the initial state $\psi$ and the Hermitian operator of the observable $A$, for notational simplicity.

The parameter-shift rule states that if the generator of the gate $G$ has only two unique eigenvalues, $e_0$ and~$e_1$, then the derivative of this circuit expectation~\eqref{function} with 
respect to the gate parameter is proportional to the difference in expectation of two circuits with shifted parameters, 
\[
\frac{d}{d\theta} f(\theta) = r \bigl[ f(\theta + \sfrac{\pi}{4r}) - f(\theta- \sfrac{\pi}{4r})\bigr] \ ,
\label{prule}
\]
where the shift constant is $r=\sfrac{a}{2}(e_1-e_0)$.
Compared to other approaches for evaluating circuit gradients~\cite{Jordan2005a}, the parameter-shift rule has the advantage that it  requires the performance of two circuits each of which is the same number of gates as the original circuit, and does not requires ancilla qubits.
We will put aside for now the practical difficulties of evaluating these expectations with sufficient accuracy on near-term NISQ (Noisy-Intermediate Scale Quantum)~\cite{Preskill2018a} computers.

Gradients of such quantum circuits are of use in the optimization step of variational quantum algorithms~\cite{McClean2016a}. In these hybrid-quantum-classical approaches, we construct a quantum circuit, and then vary the parameters to minimize some objective function of interest. Examples include the Variational Quantum Eigensolver (VQE)~\cite{Peruzzo2014a,Kandala2017a}, Quantum Approximate Optimization Algorithm (QAOA)~\cite{Farhi2014a}, quantum autoencoders~\cite{Romero2017a}, and various proposals to quantum machine learning~\cite{Schuld2018a,Schuld2019b,Havlicek2019a,Farhi2018a,Mitarai2018a}. Promising approach to the classical optimization step include gradient descent, or stochastic gradient descent if the optimization is over large input data sets, or closely related algorithms such as ADAM~\cite{Goodfellow2016a, Kingma2014a}.

The parameter-shift approach to quantum gradients can only be directly applied to gates with 2-unique eigenvalues. However, herein we will discuss how parameter-shift gradients can be evaluated for a much wider range of parameterized gates using the product rule of calculus, provided we can decompose our gate of interest into a product of gates, each of which is parameter-shift rule differentiable. We will demonstrate this idea for 2-qubit gates, and discuss several gate decompositions in detail.

In classical simulations of quantum circuits we do not need to resort to the parameter shift rule, since we can apply non-unitary operators to quantum states. We will conclude with a discussion of how to efficiently calculate gradients of quantum circuits on classical hardware.

\paragraph*{Parameter-shift rule gradients -}

Let us review why the parameter shift works. 
Suppose that  the generator of the gate $G$~\eqref{gategen} is unitary as well as Hermitian, and the prefactor $a=1$. Then $G$ is also idempotent $GG=I$, and with Euler's identity we can express the gate as
\[
U_G(\theta) = e^{-i \theta G} = I \cos( \theta) - i G \sin( \theta) \ .
\]

The key insight is that even if $G$ is not unitary, if it has only two unique eigenvalues, $e_0$ and $e_1$, then we can always convert the generator $aG$ to a unitary operator by adding and multiplying by real constants, $\sfrac{a}{r} (G - s)$, where $r=\sfrac{a}{2}(e_1-e_0)$, and $s=\sfrac{1}{2}(e_1+e_0)$. The additive shift can be neglected since it only adds an irrelevant phase. Therefore, for any real constant $a$, and Hermitian operator $G$ with two unique eigenvalues, we have
\[
\label{euler}
U_G(\theta) = e^{-i a \theta G} = 
I \cos(r \theta) - i \sfrac{a}{r} G \sin(r \theta)
\]
up to phase. And as a special case of~\eqref{euler} we have 
\[
\label{euler2}
U_G(\pm\sfrac{\pi}{4r}) = \sfrac{1}{\sqrt{2}}(I \mp i \sfrac{a}{r} G) \ .
\]

Note that the derivative of the gate~\eqref{gategen} is
\[
\label{gate_der}
\sfrac{\partial}{\partial \theta} U_G(\theta) = - ia G e^{-ia\theta G} \ .
\]
We can now derive the parameter-shift rule, Eq.~\eqref{prule}. 
\begin{subequations}
\[
&\frac{\partial}{\partial \theta} f(\theta) \notag \\
& = \langle \psi| \   [+i a G] U_G^\dagger(\theta)\   A\  U_G(\theta)\ |\psi\rangle 
 \label{a} \\
& \qquad + \langle \psi| \  U_G^\dagger(\theta)\ A\ [-i a G] U_G(\theta)\ |\psi\rangle
 \notag \\
& = \sfrac{r}{2} \langle \psi| \ U_G^\dagger(\theta) (I + i \sfrac{a}{r} G)  \ A\ (I - i \sfrac{a}{r} G)  U_G(\theta)\ |\psi\rangle 
 \label{b}  \\ 
& \qquad - \sfrac{r}{2} \langle \psi| \  U_G^\dagger(\theta) (I - i \sfrac{a}{r} G)\ A\ (I + i \sfrac{a}{r} G) U_G(\theta)\ |\psi\rangle\
 \notag\\
& = r \langle \psi| \ U_G^\dagger(\theta  + \sfrac{\pi}{4r}) \ A\ U_G(\theta + \sfrac{\pi}{4r})\ |\psi\rangle 
  \label{c} \\ 
& \qquad - r \langle \psi| \  U_G^\dagger(\theta - \sfrac{\pi}{4r}) \ A\  U_G(\theta-\sfrac{\pi}{4r})\ |\psi\rangle
\notag \\
& =   r\ \bigl[ f(\theta + \sfrac{\pi}{4r}) - f(\theta- \sfrac{\pi}{4r})\bigr]
 \label{d} 
\]
\end{subequations}
(a) We write out the gradient using~\eqref{gate_der} and the product rule; (b) and then rearrange and gather terms so that the Hermitian measurement operators are acted upon by conjugate unitary operators; (c) We recognize from \eqref{euler2} that these unitaries represent instances of the initial gate, and thus shift the gate's parameter; (d) This leads to the parameter shift rule for circuit gradients.

Note that the value of the shift constant $r$ depends upon the parameterization. For instance, the 1-qubit Pauli rotation gates are all parameter-shift differentiable with $r=\half$. 
\begin{subequations}
\[
R_X(\theta) &= e^{-i\frac{1}{2}\theta X} & r=\sfrac{1}{2} \\
R_Y(\theta) &= e^{-i\frac{1}{2}\theta Y} & r=\sfrac{1}{2} \\
R_Z(\theta) &= e^{-i\frac{1}{2}\theta Z} & r=\sfrac{1}{2}
\]
\end{subequations}
Here the Pauli matrices are $X=(\begin{smallmatrix}0 & 1 \\ 1 & 0\end{smallmatrix})$,
$Y=(\begin{smallmatrix}0 & \text{-}i \\ i & 0\end{smallmatrix})$, 
and $Z=(\begin{smallmatrix}1 & 0 \\ 0 & \text{-}1\end{smallmatrix})$.
On the other hand, it can be more convenient to represent the same gates as powers of the Pauli operators, but in that case the shift constant is $r=\sfrac{\pi}{2}$.
\begin{subequations}
\[
X^t &\simeq R_X(\pi t) = e^{-i\frac{\pi}{2}t X} & r=\sfrac{\pi}{2} \\
Y^t &\simeq R_Y(\pi t) = e^{-i\frac{\pi}{2}t Y} & r=\sfrac{\pi}{2} \\
Z^t &\simeq R_Z(\pi t) = e^{-i\frac{\pi}{2}t Z} & r=\sfrac{\pi}{2}
\]
\end{subequations}
Here we use $\simeq$ to indicate the unitaries are equal up to a phase factor.

\paragraph*{Parameter-shift gradients via gate decomposition -}
A direct application of the parameter-shift rule requires that the generator of the gate have only 2-eigenvalues. However, we can evaluate gradients for gates that do not meet this requirement 
by decomposing the dynamics into a sequence of gates, each of which has a generator of the requisite form.

As a trivial example, consider the 2-qubit canonical gate, 
\[
U_{\text{CAN}} & = \exp\Bigl(-i\sfrac{\pi}{2} (t_x \ X\otimes X + t_y Y\otimes Y + t_z\ Z\otimes Z)\Bigr) \ .
\]
This gate is of interest because it is, in a sense, the elementary 2-qubit gate. Any other 2-qubit gate can be constructed by prepending or appending local 1-qubit rotations~\cite{Zhang2003a}. 

The Hamiltonian of the canonical gate has more than~2 unique eigenvalues in general, yet we can evaluate gradients with respect to any of the three parameters using the parameter-shift rule, with $r=\sfrac{\pi}{2}$. This is because the $X\otimes X$, $Y\otimes Y$, and $Z\otimes Z$ terms in the Hamiltonian all commute, and the canonical gate can be decomposed into a sequence of  XX,  YY, and ZZ gates (in arbitrary order).
\[
\text{\small
\Qcircuit @C=0.8em @R=1em {
  & \multigate{1}{\text{CAN}(t_x,t_y, t_z)} & \qw 
  &  \dstick{=} &
  & \multigate{1}{X\!X^{t_x}} & \multigate{1}{Y\!Y^{t_y}} & \multigate{1}{Z\!Z^{t_z}}
  & \qw
  \\
  &\ghost{\text{CAN}(t_x,t_y, t_z)} & \qw & &
 & \ghost{X\!X^{t_7}} & \ghost{Y\!Y^{t_8}} & \ghost{Z\!Z^{t_9}}
& \qw
}}
\]
\[
U_{\text{XX}}(t) \simeq e^{-i\sfrac{\pi}{2} t\ X\otimes X} \notag \\
U_{\text{YY}}(t) \simeq e^{-i\sfrac{\pi}{2} t\ Y\otimes Y} \notag \\
U_{\text{ZZ}}(t) \simeq e^{-i\sfrac{\pi}{2} t\ Z\otimes Z} \notag
\]
For these parameterizations of the XX,  YY, and ZZ gates  we have $r=\sfrac{\pi}{2}$.

More generally, we can decompose any 2-qubit gate into a canonical gate plus 1-qubit gates~\cite{Zhang2003a,Zhang2004a,Blaauboer2008a,Drury2008a,Watts2013a,Peterson2019a}. 
\[
\label{canonical}
\text{\small
\Qcircuit @C=0.5em @R=1em {
  & \gate{X^{t_1}} & \gate{Y^{t_2}} & \gate{X^{t_3}} 
  & \multigate{1}{\text{CAN}(t_7, t_8, t_9)}
  & \gate{X^{t_{10}}} & \gate{Y^{t_{11}}} & \gate{X^{t_{12}}}
  & \qw
  \\
 & \gate{X^{t_{4}}} & \gate{Y^{t_{5}}} & \gate{X^{t_{6}}} 
 & \ghost{\text{CAN}(t_7, t_8, t_9)}
 & \gate{X^{t_{13}}} & \gate{Y^{t_{14}}} & \gate{X^{t_{15}}} 
& \qw
}}
\]

Provided we can determine the functional relation between the original gate parameter $\theta$ and the parameters
of the decomposition, we can evaluate gradients with the product rule. 
\[
\tfrac{d}{d\theta} f(\theta)
= \sum_{i=1,15} \tfrac{\partial }{\partial t_i} f_{\text{CAN}}(t_1, t_2, \ldots, t_{15})\ \tfrac{dt_i}{d\theta}
\]
Since an arbitrary 2-qubit gate has 15 parameters, we may need up to 30 expectation evaluations to evaluate one gradient of a 2-qubit gate.

Note that there are many essentially equivalent choices as to how to parameterize the local 1-qubit rotations in~\eqref{canonical}. 
Here, we have used the X-Y-X Euler angle decomposition, rather than the more common Z-Y-Z decomposition~\cite{Nielsen2000a}.
The canonical gate has many symmetries under local transformations, such that different coordinates 
can represent gates that  differ only by 1-qubit rotations. In particular $\operatorname{CAN}(t,0,0)$, $\operatorname{CAN}(0,t,0)$,
 and $\operatorname{CAN}(0,0,t)$ are all locally equivalent. 
 To avoid redundancy the canonical parameters can be restricted to a particular Weyl Chamber~\cite{Zhang2003a}, and traditionally this
 means that when the canonical gate has only 1 non-zero parameter we restrict to the XX-gate class, $\operatorname{CAN}(t,0,0)$. 
 By choosing an X-Y-X decomposition, the inner X-gates can be readily commuted with the XX-gate, which may simplify the decomposition.

On the other hand, when decomposing gates in the parametric swap (PSWAP) class~\cite{Smith2016a}, $\operatorname{CAN}(\sfrac{1}{2},\sfrac{1}{2}, t)$, it is advantageous to use a Z-Y-Z decomposition, since a PSWAP gate can can be decomposed as a swap, $\operatorname{CAN}(\sfrac{1}{2},\sfrac{1}{2}, \sfrac{1}{2})$, followed by a ZZ-gate, $\operatorname{CAN}(0,0, t-\sfrac{1}{2})$. A Z-gate can therefore commute past a PSWAP while switching qubits.

\paragraph*{Decomposition of the cross-resonance gate~-}
As another illustration of differentiation via gate-decomposition, we will consider a trickier example~\cite{Schuld2019a,Vidal2018a} in the cross-resonance gate family.
\[
G_\text{CR} &= X\otimes I - b\ Z\otimes X + c\ I\otimes X 
\label{CR_H}
\\
U_\text{CR} & = \exp({-i\sfrac{\pi}{2} s G_{\text{CR}}})
\]
The CR-gate is a natural gate for certain microwave-controlled transmon superconducting qubit architectures~\cite{Rigetti2010a,Chow2011a,Krantz2019a}.

\begin{figure}
\includegraphics[width=3.25in]{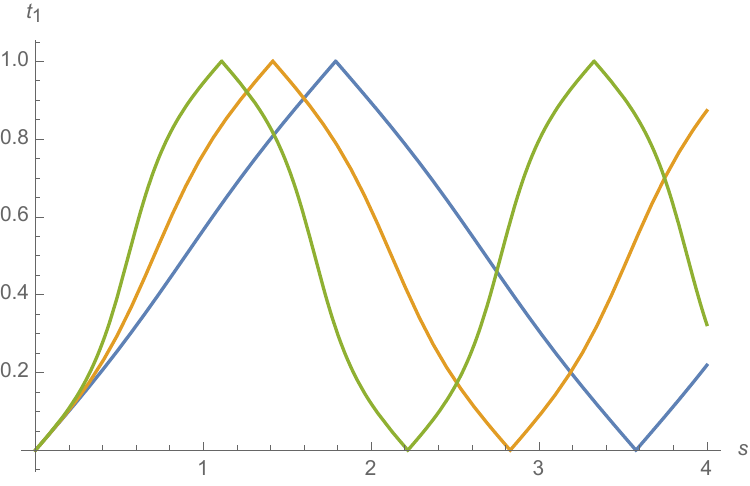}
\includegraphics[width=3.25in]{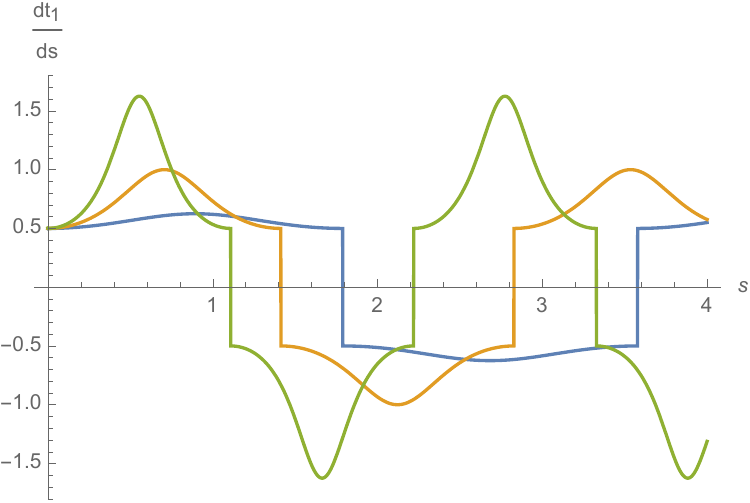}
\caption{The $t_1$ \eqref{t1} parameter, and first derivative \eqref{dt1}, of the CR canonical gate decomposition, 
for $b=\sfrac{1}{2}, 1, $ and $ \sfrac{3}{2}$ (blue, orange, green). Note that $t_1$ exhibits discontinuities, but these
are largely artifactual, since $X^{t_1}$ is the same as $X^{2-t_1}$, up to a phase factor. 
}
\end{figure}

\begin{figure}
\includegraphics[width=3.25in]{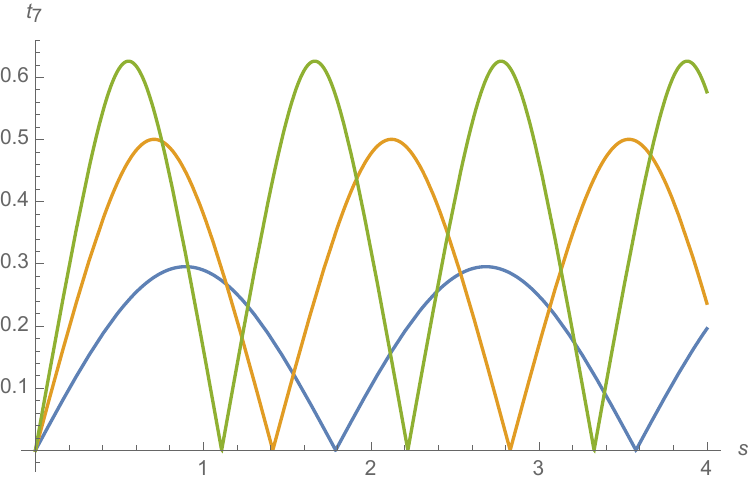}
\includegraphics[width=3.25in]{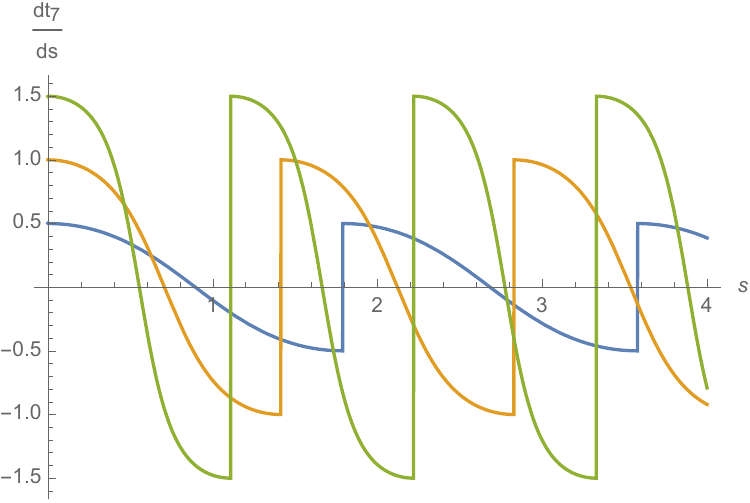}
\caption{The $t_7$ \eqref{t7} parameter, and first derivative \eqref{dt7}, of the CR canonical gate decomposition, 
for $b=\sfrac{1}{2}, 1, $ and $\sfrac{3}{2}$ (blue, orange, green). Note that to implement a CNOT gate we require $t_7=\half$, which
is achievable only for $b\geq1$. This value of $b$ is also sufficient to implement any member of the XX gate family, since $\text{CAN}(t,0,0)$ with $t\geq\half$ is locally equivalent to $\text{CAN}(\half-t,0,0)$~\cite{Zhang2003a}. Also, as $\text{CAN}(t,0,0)$ is locally equivalent to $\text{CAN}(-t,0,0)$, so (as for $t_1$) the discontinuities are somewhat artifactual.
}
\end{figure}

This CR-gate can be represented by the following circuit diagram.
\[
\text{\small\small
\Qcircuit @C=0.5em @R=1.5em {
& \qw& \multigate{1}{\text{CR}(s;b,c)}& \qw  & \qw \\
& \qw& \ghost{\text{CR}(s;b,c)}& \qw & \qw
}}
\]
which can be canonically decomposed as a circuit of 1-qubit gates and an XX-gate.
\[
\label{cr_circ}
\text{\small
\Qcircuit @C=0.5em @R=1em {
  & \gate{X^{t_1}} & \gate{Y^{\frac{3}{2}}} & \gate{X} & \multigate{1}{X\!X^{t_7}} & \gate{Y^{\frac{3}{2}}} & \gate{X} & \gate{X^{t_1}} & \qw \\
 & \gate{X^{t_4}} & \qw & \qw & \ghost{X\!X^{t_7}} & \qw & \qw & \qw  & \qw
}}
\]
An equivalent decomposition, using CNOTs as the 2-qubit interaction, is:
\[
\text{\small \Qcircuit @C=0.5em @R=1em {
 & \gate{X^{t_1}} & \gate{Y^{\frac{3}{2}}} & \gate{X} & \gate{H} & \ctrl{1} & \qw & \ctrl{1} & \gate{H} & \gate{Y^{\frac{3}{2}}} & \gate{X} & \gate{X^{t_1}} & \qw \\
& \gate{X^{t_4}} & \qw & \qw & \gate{H} & \targ & \gate{Z^{t_7}} & \targ & \gate{H} & \qw & \qw & \qw & \qw
}}
\]
The 3 non-trivial parameters of this circuit can be expressed using elementary functions of the CR-gate parameters.
\[
\label{t1}
t_1 & =  \sfrac{1}{\pi} \arccos\left(\frac{\cos(\sfrac{\pi}{2} \sqrt{1 + b^2}s) }{ \cos(\sfrac{\pi}{2}t_7)}\right)   \\
\label{t4}
t_4 & = c s \\ 
\label{t7}
t_7 & = \sfrac{1}{\pi} \arccos\left(\frac{1 + b^2  \cos(\pi \sqrt{1 + b^2} s)}{ 1 + b^2}\right)
\]

The procedure to decompose the CR-gate is as follows:
We first guess the circuit ansatz~\eqref{cr_circ} by examining numerical decompositions of the CR-gate~\cite{QuantumFlowGradients}. We are left with 3 undetermined parameters. Since the CR-gate is equivalent to an XX-gate up to 1-qubit rotations, the $t_7$ parameter can be derived using Eq.~25 of Ref.~\cite{Zhang2003a}.  
\[
t_7 = \frac{1}{\pi} \arccos\Left(\tfrac{1}{4} \text{Tr}\Left[ (\mathcal{M}^\dagger U_\text{CR} \mathcal{M})^T (\mathcal{M}^\dagger U_\text{CR} \mathcal{M})\Right]\Right)
\]
Here $\mathcal{M}$ is the magic gate, which transforms to a magic basis~\cite{Makhlin2002a,Vatan2004a}.
(In a magic basis the Kronecker products of two 1-qubit gates are orthogonal matrices.)
\[
\mathcal{M} = \frac{1}{\sqrt{2}} 
\begin{pmatrix} 
1 & 0 & 0 & i \\
0 & i & 1 & 0 \\ 
0 & i &-1 & 0 \\
1 & 0 & 0 & -i 
\end{pmatrix}
\]
\[
\notag
\Qcircuit @C=1.0em @R=1em {
& \multigate{1}{{\mathcal{M}}} & \qw &  \dstick{=} & & \gate{S} &\qw & \targ & \qw \\
& \ghost{\mathcal{M}} &\qw & & & \gate{S} & \gate{H} & \ctrl{-1}& \qw
}
\]
The $I\otimes X$ term in the Hamiltonian commutes with the other two terms, and can be separated out as an X rotation on the second qubit, which gives the $t_4$ parameter. The final parameter, $t_1$,  can then be solved for analytically.

We can therefore calculate gradients of the CR-gate using the product rule, 8 expectation evaluations using the parameter-shift rule, and the following derivatives.
\[
\label{dt1}
\frac{dt_1}{ds} & =
\frac{
	\sqrt{1 + b^2} \sec(\sfrac{\pi}{2} t_7)
	\sin(\sqrt{1 + b^2} \sfrac{\pi}{2} s)
}{
	2 \sqrt{
		1 - \cos^2(\sqrt{1 + b^2} \sfrac{\pi}{2} s) \sec^2(\sfrac{\pi}{2} t_7)
		} 
}
\ \ 
\frac{dt_7}{ds}
 \\
\frac{dt_4}{ds} &=c \\
\label{dt7}
\frac{dt_7}{ds} &= \frac{b^2 \sin(\pi \sqrt{b^2 + 1} s)}{\sqrt{b^2 + 1} \sqrt{ \frac{1 - (b^2 \cos(\pi \sqrt{b^2 + 1)} s) + 1)^2}{(b^2 + 1)^2}}}
\]

\paragraph*{Binary decomposition of the cross-resonance gate -}
As it happens, we do not need to resort to a full decomposition of the CR-gate to evaluate the gradient.
The CR-Hamiltonian~\eqref{CR_H} has 4 unique eigenvalues in general, $\pm c \pm \sqrt{b^2+1}$~\cite{Vidal2018a}.
 However, if $c$ is zero there are only two unique eigenvalues, and the parameter-shift rule applies.
Since the $I\otimes X$ component of the Hamiltonian~\eqref{CR_H} commutes, we can separate out the $c$ parameter onto a separate $X$ rotation of the second qubit, as we did for the full canonical decomposition. 
Thus we can decompose the full CR-gate into just two components, each of which is parameter-shift rule differentiable with respect to $s$.
\[
\text{\small
\Qcircuit @C=1.0em @R=1em {
& \multigate{1}{\text{CR}(s;b,c)}& \qw  &   \dstick{=} &  &\multigate{1}{\text{CR}(s;b,0)}& \qw  & \qw \\
& \ghost{\text{CR}(s;b,c)}& \qw & &  & \ghost{\text{CR}(s;b,0)}& \gate{X^{cs}} &\qw
}}
\]
The shift  parameters for the this circuit are $r=\sfrac{\pi}{2}\sqrt{b^2+1}$ and $r=\sfrac{\pi c}{2}$ respectively.

\paragraph*{Middle-out quantum gradients on classical hardware -}
In a classical simulation of a quantum computer we can apply arbitrary operators to quantum states, and therefore we can efficiently calculate gradients of quantum circuits without resorting to the parameter-shift rule. We could back-propagate the gradients using the chain rule~\cite{Rumelhart1986a, Goodfellow2016a, Leung2017a, Tamayo-Mendoza2018a, QuantumFlowGradients}. However, this requires storing the intermediate states during the forward  propagation, resulting in a memory demand that scales as $O(2^M N)$ for  $M$ qubits and $N$ gates.  Fortunately, we can simplify the procedure and reduce the memory requirements by taking advantage of the time-reversibility of quantum mechanics~\cite{Khaneja2005a,Leung2017a}.

Suppose we have a quantum circuit composed of $N$ parameterized gates.
\[
U(\boldsymbol{\theta}) = U_N(\theta_N) \ldots U_k(\theta_k) \ldots U_2(\theta_2)\ U_1(\theta_1)
\]
Then the derivative of our observable~\eqref{function} with respect to one of the parameters is
\[
\frac{d f(\boldsymbol{\theta})}{d\theta_k} & = \langle\psi| U_1^\dagger(\theta_1) \ldots U_k^\dagger(\theta_k)  \ldots U_N^\dagger(\theta_N) \cdot A \cdot \notag\\
 & \qquad   U_N(\theta_N) \ldots [-i a_k G_k] U_k(\theta_k) \ldots U_1(\theta_1)  |\psi\rangle
\notag  \\ & \qquad\qquad+ h.c.
\]
Here $a_k$ and $G_k$ are the scaling constant and Hermitian generator of the $k$th gate~\eqref{gategen}.

We can rewrite this expression in a more compact form.
\[
\frac{d f(\boldsymbol{\theta})}{d\theta_k} &= -2 a_k \Im \langle B_{k}|\ G_k\ | F_{k} \rangle
 \\
|F_{k} \rangle &= U_k(\theta_k) \ldots U_2(\theta_2) U_1(\theta_1)  |\psi\rangle \notag
\\
| B_{k} \rangle &= U^\dagger_{k+1}(\theta_{k+1})  \ldots U^\dagger_N(\theta_N)  \cdot A \cdot 
\notag \\ & \qquad\qquad U_N(\theta_N) \ldots U_2(\theta_2) U_1(\theta_1)  |\psi\rangle \notag
\]
Here $|F_{k} \rangle$ is the initial state propagated forward in time up to the $k$th gate, and $| B_{k} \rangle$ is the initial state propagated forward in time through the entire circuit, followed by an application of the Hermitian observable, followed by a reversed time propagation backwards to the $(k+1)$th gate. Note that  $| B_{k} \rangle$  is not normalized due to the application of the Hermitian operator. 

The trick is that if we are evaluating all of the gradients, we can recursively evaluate the forward and backward states, which requires only one additional gate application each per step.
\[
|F_{k+1} \rangle  & = U_{k+1}(\theta_{k+1}) |F_{k} \rangle 
\\
|B_{k+1} \rangle  & = U_{k+1}(\theta_{k+1}) |B_{k} \rangle
\notag
\]

Evaluating the original circuit of $N$ gates requires storage of one state, $N$ gate applications, and one inner product evaluation of the observable. Evaluating 
all $N$ gradients using the middle-out approach requires $4N$ gate evaluations, $N$ applications of gate generators, and $N$ inner products. Thus evaluating all the gradients requires only about twice the memory, and around $6$ times the computational time needed to evaluate the original circuit.

\paragraph*{Acknowledgments}
We would like to thank Eric C.\ Peterson for insightful discussions.
This research undertaken while a guest of Prof.\ Garnet Chan at CalTech.
Support provided by US DOE, Office of Science, Grant No.\ DE-SC0019374.


\bibliography{Quantum}

\begin{thebibliography}{36}
\expandafter\ifx\csname natexlab\endcsname\relax\def\natexlab#1{#1}\fi
\expandafter\ifx\csname bibnamefont\endcsname\relax
  \def\bibnamefont#1{#1}\fi
\expandafter\ifx\csname bibfnamefont\endcsname\relax
  \def\bibfnamefont#1{#1}\fi
\expandafter\ifx\csname citenamefont\endcsname\relax
  \def\citenamefont#1{#1}\fi
\expandafter\ifx\csname url\endcsname\relax
  \def\url#1{\texttt{#1}}\fi
\expandafter\ifx\csname urlprefix\endcsname\relax\def\urlprefix{URL }\fi
\providecommand{\bibinfo}[2]{#2}
\providecommand{\eprint}[2][]{\url{#2}}

\bibitem[{\citenamefont{Mitarai et~al.}(2018)\citenamefont{Mitarai, Negoro,
  Kitagawa, and Fujii}}]{Mitarai2018a}
\bibinfo{author}{\bibfnamefont{K.}~\bibnamefont{Mitarai}},
  \bibinfo{author}{\bibfnamefont{M.}~\bibnamefont{Negoro}},
  \bibinfo{author}{\bibfnamefont{M.}~\bibnamefont{Kitagawa}}, \bibnamefont{and}
  \bibinfo{author}{\bibfnamefont{K.}~\bibnamefont{Fujii}},
  \bibinfo{journal}{Phys. Rev. A} \textbf{\bibinfo{volume}{98}},
  \bibinfo{pages}{032309} (\bibinfo{year}{2018}),
  \bibinfo{note}{{arXiv:1803.00745}}.

\bibitem[{\citenamefont{Li et~al.}(2017)\citenamefont{Li, Yang, Peng, and
  Sun}}]{Li2017a}
\bibinfo{author}{\bibfnamefont{J.}~\bibnamefont{Li}},
  \bibinfo{author}{\bibfnamefont{X.}~\bibnamefont{Yang}},
  \bibinfo{author}{\bibfnamefont{X.}~\bibnamefont{Peng}}, \bibnamefont{and}
  \bibinfo{author}{\bibfnamefont{C.-P.} \bibnamefont{Sun}},
  \bibinfo{journal}{Phys. Rev. Lett.} \textbf{\bibinfo{volume}{118}},
  \bibinfo{pages}{150503} (\bibinfo{year}{2017}).

\bibitem[{\citenamefont{Schuld et~al.}(2019)\citenamefont{Schuld, Bergholm,
  Gogolin, Izaac, and Killoran}}]{Schuld2019a}
\bibinfo{author}{\bibfnamefont{M.}~\bibnamefont{Schuld}},
  \bibinfo{author}{\bibfnamefont{V.}~\bibnamefont{Bergholm}},
  \bibinfo{author}{\bibfnamefont{C.}~\bibnamefont{Gogolin}},
  \bibinfo{author}{\bibfnamefont{J.}~\bibnamefont{Izaac}}, \bibnamefont{and}
  \bibinfo{author}{\bibfnamefont{N.}~\bibnamefont{Killoran}},
  \bibinfo{journal}{Phys. Rev. A} \textbf{\bibinfo{volume}{99}},
  \bibinfo{pages}{032331} (\bibinfo{year}{2019}),
  \bibinfo{note}{{arXiv:1811.11184}}.

\bibitem[{\citenamefont{Bergholm et~al.}(2018)\citenamefont{Bergholm, Izaac,
  Schuld, Gogolin, and Killoran}}]{Bergholm2018a}
\bibinfo{author}{\bibfnamefont{V.}~\bibnamefont{Bergholm}},
  \bibinfo{author}{\bibfnamefont{J.}~\bibnamefont{Izaac}},
  \bibinfo{author}{\bibfnamefont{M.}~\bibnamefont{Schuld}},
  \bibinfo{author}{\bibfnamefont{C.}~\bibnamefont{Gogolin}}, \bibnamefont{and}
  \bibinfo{author}{\bibfnamefont{N.}~\bibnamefont{Killoran}}
  (\bibinfo{year}{2018}), \bibinfo{note}{{arXiv:1811.04968}}.

\bibitem[{\citenamefont{Vidal and Theis}(2018)}]{Vidal2018a}
\bibinfo{author}{\bibfnamefont{J.~G.} \bibnamefont{Vidal}} \bibnamefont{and}
  \bibinfo{author}{\bibfnamefont{D.~O.} \bibnamefont{Theis}}
  (\bibinfo{year}{2018}), \bibinfo{note}{{arXiv:1812.06323}}.

\bibitem[{\citenamefont{Jordan}(2005)}]{Jordan2005a}
\bibinfo{author}{\bibfnamefont{S.~P.} \bibnamefont{Jordan}},
  \bibinfo{journal}{Phys. Rev. Lett.} \textbf{\bibinfo{volume}{95}},
  \bibinfo{pages}{050501} (\bibinfo{year}{2005}).

\bibitem[{\citenamefont{Preskill}(2018)}]{Preskill2018a}
\bibinfo{author}{\bibfnamefont{J.}~\bibnamefont{Preskill}},
  \bibinfo{journal}{Quantum} \textbf{\bibinfo{volume}{2}}, \bibinfo{pages}{79}
  (\bibinfo{year}{2018}), \bibinfo{note}{arXiv:1801.00862}.

\bibitem[{\citenamefont{McClean et~al.}(2016)\citenamefont{McClean, Romero,
  Babbush, and Aspuru-Guzik}}]{McClean2016a}
\bibinfo{author}{\bibfnamefont{J.~R.} \bibnamefont{McClean}},
  \bibinfo{author}{\bibfnamefont{J.}~\bibnamefont{Romero}},
  \bibinfo{author}{\bibfnamefont{R.}~\bibnamefont{Babbush}}, \bibnamefont{and}
  \bibinfo{author}{\bibfnamefont{A.}~\bibnamefont{Aspuru-Guzik}},
  \bibinfo{journal}{New J. Phys} \textbf{\bibinfo{volume}{18}},
  \bibinfo{pages}{023023} (\bibinfo{year}{2016}).

\bibitem[{\citenamefont{Peruzzo et~al.}(2014)\citenamefont{Peruzzo, McClean,
  Shadbolt, Yung, Zhou, Love, Aspuru-Guzik, and O'Brien}}]{Peruzzo2014a}
\bibinfo{author}{\bibfnamefont{A.}~\bibnamefont{Peruzzo}},
  \bibinfo{author}{\bibfnamefont{J.}~\bibnamefont{McClean}},
  \bibinfo{author}{\bibfnamefont{P.}~\bibnamefont{Shadbolt}},
  \bibinfo{author}{\bibfnamefont{M.-H.} \bibnamefont{Yung}},
  \bibinfo{author}{\bibfnamefont{X.-Q.} \bibnamefont{Zhou}},
  \bibinfo{author}{\bibfnamefont{P.~J.} \bibnamefont{Love}},
  \bibinfo{author}{\bibfnamefont{A.}~\bibnamefont{Aspuru-Guzik}},
  \bibnamefont{and} \bibinfo{author}{\bibfnamefont{J.~L.}
  \bibnamefont{O'Brien}}, \bibinfo{journal}{Nat. Commun}
  \textbf{\bibinfo{volume}{5}}, \bibinfo{pages}{4213} (\bibinfo{year}{2014}).

\bibitem[{\citenamefont{Kandala et~al.}(2017)\citenamefont{Kandala, Mezzacapo,
  Temme, Takita, Brink, Chow, and Gambetta}}]{Kandala2017a}
\bibinfo{author}{\bibfnamefont{A.}~\bibnamefont{Kandala}},
  \bibinfo{author}{\bibfnamefont{A.}~\bibnamefont{Mezzacapo}},
  \bibinfo{author}{\bibfnamefont{K.}~\bibnamefont{Temme}},
  \bibinfo{author}{\bibfnamefont{M.}~\bibnamefont{Takita}},
  \bibinfo{author}{\bibfnamefont{M.}~\bibnamefont{Brink}},
  \bibinfo{author}{\bibfnamefont{J.~M.} \bibnamefont{Chow}}, \bibnamefont{and}
  \bibinfo{author}{\bibfnamefont{J.~M.} \bibnamefont{Gambetta}},
  \bibinfo{journal}{Nature} \textbf{\bibinfo{volume}{549}},
  \bibinfo{pages}{242} (\bibinfo{year}{2017}),
  \bibinfo{note}{arXiv:1704.05018}.

\bibitem[{\citenamefont{Farhi et~al.}(2014)\citenamefont{Farhi, Goldstone, and
  Gutmann}}]{Farhi2014a}
\bibinfo{author}{\bibfnamefont{E.}~\bibnamefont{Farhi}},
  \bibinfo{author}{\bibfnamefont{J.}~\bibnamefont{Goldstone}},
  \bibnamefont{and} \bibinfo{author}{\bibfnamefont{S.}~\bibnamefont{Gutmann}}
  (\bibinfo{year}{2014}), \bibinfo{note}{arXiv:1411.4028}.

\bibitem[{\citenamefont{Romero et~al.}(2017)\citenamefont{Romero, Olson, and
  Aspuru-Guzik}}]{Romero2017a}
\bibinfo{author}{\bibfnamefont{J.}~\bibnamefont{Romero}},
  \bibinfo{author}{\bibfnamefont{J.~P.} \bibnamefont{Olson}}, \bibnamefont{and}
  \bibinfo{author}{\bibfnamefont{A.}~\bibnamefont{Aspuru-Guzik}},
  \bibinfo{journal}{Quantum Sci. Tech.} \textbf{\bibinfo{volume}{2}},
  \bibinfo{pages}{045001} (\bibinfo{year}{2017}).

\bibitem[{\citenamefont{Schuld et~al.}(2018)\citenamefont{Schuld, Bocharov,
  Svore, and Wiebe}}]{Schuld2018a}
\bibinfo{author}{\bibfnamefont{M.}~\bibnamefont{Schuld}},
  \bibinfo{author}{\bibfnamefont{A.}~\bibnamefont{Bocharov}},
  \bibinfo{author}{\bibfnamefont{K.}~\bibnamefont{Svore}}, \bibnamefont{and}
  \bibinfo{author}{\bibfnamefont{N.}~\bibnamefont{Wiebe}}
  (\bibinfo{year}{2018}), \bibinfo{note}{{arXiv:1804.00633}}.

\bibitem[{\citenamefont{Schuld and Killoran}(2019)}]{Schuld2019b}
\bibinfo{author}{\bibfnamefont{M.}~\bibnamefont{Schuld}} \bibnamefont{and}
  \bibinfo{author}{\bibfnamefont{N.}~\bibnamefont{Killoran}},
  \bibinfo{journal}{Phys. Rev. Lett.} \textbf{\bibinfo{volume}{122}},
  \bibinfo{pages}{040504} (\bibinfo{year}{2019}),
  \bibinfo{note}{{arXiv:1803.07128}}.

\bibitem[{\citenamefont{Havlicek et~al.}(2019)\citenamefont{Havlicek,
  C{\'o}rcoles, Temme, Harrow, Chow, and Gambetta}}]{Havlicek2019a}
\bibinfo{author}{\bibfnamefont{V.}~\bibnamefont{Havlicek}},
  \bibinfo{author}{\bibfnamefont{A.~D.} \bibnamefont{C{\'o}rcoles}},
  \bibinfo{author}{\bibfnamefont{K.}~\bibnamefont{Temme}},
  \bibinfo{author}{\bibfnamefont{A.~W.} \bibnamefont{Harrow}},
  \bibinfo{author}{\bibfnamefont{J.~M.} \bibnamefont{Chow}}, \bibnamefont{and}
  \bibinfo{author}{\bibfnamefont{J.~M.} \bibnamefont{Gambetta}},
  \bibinfo{journal}{Nature} \textbf{\bibinfo{volume}{567}},
  \bibinfo{pages}{209} (\bibinfo{year}{2019}),
  \bibinfo{note}{arXiv:1804.11326}.

\bibitem[{\citenamefont{Farhi and Neven}(2018)}]{Farhi2018a}
\bibinfo{author}{\bibfnamefont{E.}~\bibnamefont{Farhi}} \bibnamefont{and}
  \bibinfo{author}{\bibfnamefont{H.}~\bibnamefont{Neven}}
  (\bibinfo{year}{2018}), \bibinfo{note}{{arXiv:1802.06002}}.

\bibitem[{\citenamefont{Goodfellow et~al.}(2016)\citenamefont{Goodfellow,
  Bengio, and Courville}}]{Goodfellow2016a}
\bibinfo{author}{\bibfnamefont{I.}~\bibnamefont{Goodfellow}},
  \bibinfo{author}{\bibfnamefont{Y.}~\bibnamefont{Bengio}}, \bibnamefont{and}
  \bibinfo{author}{\bibfnamefont{A.}~\bibnamefont{Courville}},
  \emph{\bibinfo{title}{Deep Learning}} (\bibinfo{publisher}{MIT Press},
  \bibinfo{address}{Cambridge, Massachusetts}, \bibinfo{year}{2016}),
  \urlprefix\url{http://www.deeplearningbook.org/}.

\bibitem[{\citenamefont{Kingma and Ba}(2014)}]{Kingma2014a}
\bibinfo{author}{\bibfnamefont{D.~P.} \bibnamefont{Kingma}} \bibnamefont{and}
  \bibinfo{author}{\bibfnamefont{J.}~\bibnamefont{Ba}} (\bibinfo{year}{2014}),
  \bibinfo{note}{arXiv:1412.6980}.

\bibitem[{\citenamefont{Zhang et~al.}(2003)\citenamefont{Zhang, Vala, Sastry,
  and Whaley}}]{Zhang2003a}
\bibinfo{author}{\bibfnamefont{J.}~\bibnamefont{Zhang}},
  \bibinfo{author}{\bibfnamefont{J.}~\bibnamefont{Vala}},
  \bibinfo{author}{\bibfnamefont{S.}~\bibnamefont{Sastry}}, \bibnamefont{and}
  \bibinfo{author}{\bibfnamefont{K.~B.} \bibnamefont{Whaley}},
  \bibinfo{journal}{Phys. Rev. A} \textbf{\bibinfo{volume}{67}},
  \bibinfo{pages}{042313} (\bibinfo{year}{2003}),
  \bibinfo{note}{{arXiv:quant-ph/0209120}}.

\bibitem[{\citenamefont{Zhang et~al.}(2004)\citenamefont{Zhang, Vala, Sastry,
  and Whaley}}]{Zhang2004a}
\bibinfo{author}{\bibfnamefont{J.}~\bibnamefont{Zhang}},
  \bibinfo{author}{\bibfnamefont{J.}~\bibnamefont{Vala}},
  \bibinfo{author}{\bibfnamefont{S.}~\bibnamefont{Sastry}}, \bibnamefont{and}
  \bibinfo{author}{\bibfnamefont{K.~B.} \bibnamefont{Whaley}},
  \bibinfo{journal}{Phys. Rev. A} \textbf{\bibinfo{volume}{69}},
  \bibinfo{pages}{042309} (\bibinfo{year}{2004}),
  \bibinfo{note}{arXiv:quant-ph/0308167}.

\bibitem[{\citenamefont{Blaauboer and de~Visser}(2008)}]{Blaauboer2008a}
\bibinfo{author}{\bibfnamefont{M.}~\bibnamefont{Blaauboer}} \bibnamefont{and}
  \bibinfo{author}{\bibfnamefont{R.~L.} \bibnamefont{de~Visser}},
  \bibinfo{journal}{J. Phys. A : Math. Theor} \textbf{\bibinfo{volume}{41}},
  \bibinfo{pages}{395307} (\bibinfo{year}{2008}),
  \bibinfo{note}{{arXiv:cond-mat/0609750}}.

\bibitem[{\citenamefont{Drury and Love}(2008)}]{Drury2008a}
\bibinfo{author}{\bibfnamefont{B.}~\bibnamefont{Drury}} \bibnamefont{and}
  \bibinfo{author}{\bibfnamefont{P.}~\bibnamefont{Love}}, \bibinfo{journal}{J.
  Phys. A : Math. Theor} \textbf{\bibinfo{volume}{41}}, \bibinfo{pages}{395305}
  (\bibinfo{year}{2008}), \bibinfo{note}{{arXiv:0806.4015}}.

\bibitem[{\citenamefont{Watts et~al.}(2013)\citenamefont{Watts, O'Connor, and
  Vala}}]{Watts2013a}
\bibinfo{author}{\bibfnamefont{P.}~\bibnamefont{Watts}},
  \bibinfo{author}{\bibfnamefont{M.}~\bibnamefont{O'Connor}}, \bibnamefont{and}
  \bibinfo{author}{\bibfnamefont{J.}~\bibnamefont{Vala}},
  \bibinfo{journal}{Entropy} \textbf{\bibinfo{volume}{15}},
  \bibinfo{pages}{1963} (\bibinfo{year}{2013}).

\bibitem[{\citenamefont{Peterson et~al.}(2019)\citenamefont{Peterson, Crooks,
  and Smith}}]{Peterson2019a}
\bibinfo{author}{\bibfnamefont{E.~C.} \bibnamefont{Peterson}},
  \bibinfo{author}{\bibfnamefont{G.~E.} \bibnamefont{Crooks}},
  \bibnamefont{and} \bibinfo{author}{\bibfnamefont{R.~S.} \bibnamefont{Smith}}
  (\bibinfo{year}{2019}), \bibinfo{note}{\arxiv{1904.10541}}.

\bibitem[{\citenamefont{Nielsen and Chuang}(2000)}]{Nielsen2000a}
\bibinfo{author}{\bibfnamefont{M.~A.} \bibnamefont{Nielsen}} \bibnamefont{and}
  \bibinfo{author}{\bibfnamefont{I.~L.} \bibnamefont{Chuang}},
  \emph{\bibinfo{title}{Quantum Computation and Quantum Information}}
  (\bibinfo{publisher}{Cambridge University Press}, \bibinfo{year}{2000}).

\bibitem[{\citenamefont{Smith et~al.}(2016)\citenamefont{Smith, Curtis, and
  Zeng}}]{Smith2016a}
\bibinfo{author}{\bibfnamefont{R.~S.} \bibnamefont{Smith}},
  \bibinfo{author}{\bibfnamefont{M.~J.} \bibnamefont{Curtis}},
  \bibnamefont{and} \bibinfo{author}{\bibfnamefont{W.~J.} \bibnamefont{Zeng}}
  (\bibinfo{year}{2016}), \bibinfo{note}{arXiv:1608.03355}.

\bibitem[{\citenamefont{Rigetti and Devoret}(2010)}]{Rigetti2010a}
\bibinfo{author}{\bibfnamefont{C.}~\bibnamefont{Rigetti}} \bibnamefont{and}
  \bibinfo{author}{\bibfnamefont{M.}~\bibnamefont{Devoret}},
  \bibinfo{journal}{Phys. Rev. B} \textbf{\bibinfo{volume}{81}},
  \bibinfo{pages}{134507} (\bibinfo{year}{2010}).

\bibitem[{\citenamefont{Chow et~al.}(2011)\citenamefont{Chow, C\'orcoles,
  Gambetta, Rigetti, Johnson, Smolin, Rozen, Keefe, Rothwell, Ketchen
  et~al.}}]{Chow2011a}
\bibinfo{author}{\bibfnamefont{J.~M.} \bibnamefont{Chow}},
  \bibinfo{author}{\bibfnamefont{A.~D.} \bibnamefont{C\'orcoles}},
  \bibinfo{author}{\bibfnamefont{J.~M.} \bibnamefont{Gambetta}},
  \bibinfo{author}{\bibfnamefont{C.}~\bibnamefont{Rigetti}},
  \bibinfo{author}{\bibfnamefont{B.~R.} \bibnamefont{Johnson}},
  \bibinfo{author}{\bibfnamefont{J.~A.} \bibnamefont{Smolin}},
  \bibinfo{author}{\bibfnamefont{J.~R.} \bibnamefont{Rozen}},
  \bibinfo{author}{\bibfnamefont{G.~A.} \bibnamefont{Keefe}},
  \bibinfo{author}{\bibfnamefont{M.~B.} \bibnamefont{Rothwell}},
  \bibinfo{author}{\bibfnamefont{M.~B.} \bibnamefont{Ketchen}},
  \bibnamefont{et~al.}, \bibinfo{journal}{Phys. Rev. Lett.}
  \textbf{\bibinfo{volume}{107}}, \bibinfo{pages}{080502}
  (\bibinfo{year}{2011}).

\bibitem[{\citenamefont{Krantz et~al.}(2019)\citenamefont{Krantz, Kjaergaard,
  Yan, Orlando, Gustavsson, and Oliver}}]{Krantz2019a}
\bibinfo{author}{\bibfnamefont{P.}~\bibnamefont{Krantz}},
  \bibinfo{author}{\bibfnamefont{M.}~\bibnamefont{Kjaergaard}},
  \bibinfo{author}{\bibfnamefont{F.}~\bibnamefont{Yan}},
  \bibinfo{author}{\bibfnamefont{T.~P.} \bibnamefont{Orlando}},
  \bibinfo{author}{\bibfnamefont{S.}~\bibnamefont{Gustavsson}},
  \bibnamefont{and} \bibinfo{author}{\bibfnamefont{W.~D.} \bibnamefont{Oliver}}
  (\bibinfo{year}{2019}), \bibinfo{note}{arXiv:1904.06560}.

\bibitem[{\citenamefont{Crooks}()}]{QuantumFlowGradients}
\bibinfo{author}{\bibfnamefont{G.~E.} \bibnamefont{Crooks}},
  \emph{\bibinfo{title}{{QuantumFlow: A Quantum Algorithms Development Toolkit}
  v0.9}}, \bibinfo{note}{\url{https://quantumflow.readthedocs.io/} See the {\tt
  decompositions} subpackage for a python implementation of canonical gate
  decomposition; and the {\tt gradients} subpackage for implementations of
  parameter-shift and middle-out gradients of parameterized quantum circuits.}

\bibitem[{\citenamefont{Makhlin}(2002)}]{Makhlin2002a}
\bibinfo{author}{\bibfnamefont{Y.}~\bibnamefont{Makhlin}},
  \bibinfo{journal}{Quant. Info. Processing} \textbf{\bibinfo{volume}{1}},
  \bibinfo{pages}{243} (\bibinfo{year}{2002}).

\bibitem[{\citenamefont{Vatan and Williams}(2004)}]{Vatan2004a}
\bibinfo{author}{\bibfnamefont{F.}~\bibnamefont{Vatan}} \bibnamefont{and}
  \bibinfo{author}{\bibfnamefont{C.}~\bibnamefont{Williams}},
  \bibinfo{journal}{Phys. Rev. A} \textbf{\bibinfo{volume}{69}},
  \bibinfo{pages}{032315} (\bibinfo{year}{2004}),
  \bibinfo{note}{arXiv:quant-ph/0308006}.

\bibitem[{\citenamefont{Rumelhart et~al.}(1986)\citenamefont{Rumelhart, Hinton,
  and Williams}}]{Rumelhart1986a}
\bibinfo{author}{\bibfnamefont{D.~E.} \bibnamefont{Rumelhart}},
  \bibinfo{author}{\bibfnamefont{G.~E.} \bibnamefont{Hinton}},
  \bibnamefont{and} \bibinfo{author}{\bibfnamefont{R.~J.}
  \bibnamefont{Williams}}, \bibinfo{journal}{Nature}
  \textbf{\bibinfo{volume}{323}}, \bibinfo{pages}{533} (\bibinfo{year}{1986}).

\bibitem[{\citenamefont{Leung et~al.}(2017)\citenamefont{Leung, Abdelhafez,
  Koch, and Schuster}}]{Leung2017a}
\bibinfo{author}{\bibfnamefont{N.}~\bibnamefont{Leung}},
  \bibinfo{author}{\bibfnamefont{M.}~\bibnamefont{Abdelhafez}},
  \bibinfo{author}{\bibfnamefont{J.}~\bibnamefont{Koch}}, \bibnamefont{and}
  \bibinfo{author}{\bibfnamefont{D.}~\bibnamefont{Schuster}},
  \bibinfo{journal}{Phys. Rev. A} \textbf{\bibinfo{volume}{95}},
  \bibinfo{pages}{042318} (\bibinfo{year}{2017}).

\bibitem[{\citenamefont{Tamayo-Mendoza
  et~al.}(2018)\citenamefont{Tamayo-Mendoza, Kreisbeck, Lindh, and
  Aspuru-Guzik}}]{Tamayo-Mendoza2018a}
\bibinfo{author}{\bibfnamefont{T.}~\bibnamefont{Tamayo-Mendoza}},
  \bibinfo{author}{\bibfnamefont{C.}~\bibnamefont{Kreisbeck}},
  \bibinfo{author}{\bibfnamefont{R.}~\bibnamefont{Lindh}}, \bibnamefont{and}
  \bibinfo{author}{\bibfnamefont{A.}~\bibnamefont{Aspuru-Guzik}},
  \bibinfo{journal}{ACS Cent. Sci.} \textbf{\bibinfo{volume}{4}},
  \bibinfo{pages}{559} (\bibinfo{year}{2018}),
  \bibinfo{note}{{arXiv:1711.08127}}.

\bibitem[{\citenamefont{Khaneja et~al.}(2005)\citenamefont{Khaneja, Reiss,
  Kehlet, Schulte-Herbr{\"u}ggen, and Glaser}}]{Khaneja2005a}
\bibinfo{author}{\bibfnamefont{N.}~\bibnamefont{Khaneja}},
  \bibinfo{author}{\bibfnamefont{T.}~\bibnamefont{Reiss}},
  \bibinfo{author}{\bibfnamefont{C.}~\bibnamefont{Kehlet}},
  \bibinfo{author}{\bibfnamefont{T.}~\bibnamefont{Schulte-Herbr{\"u}ggen}},
  \bibnamefont{and} \bibinfo{author}{\bibfnamefont{S.~J.}
  \bibnamefont{Glaser}}, \bibinfo{journal}{J. Magn. Reson}
  \textbf{\bibinfo{volume}{172}}, \bibinfo{pages}{296} (\bibinfo{year}{2005}).

\end{thebibliography}

\onecolumngrid  

\end{document}